\documentstyle[preprint,aps]{revtex}

\begin{document}
\draft
\preprint{IFT/14/93}

\title   {
           Particle-drip lines from the Hartree-Fock-Bogoliubov
                         theory with Skyrme interaction
         }
\author  {                      R. Smola\'nczuk
         }
\address {
                     So{\l}tan Institute for Nuclear Studies,
                         Ho\.za 69, 00-681 Warsaw, Poland
         }
\author  {                      J. Dobaczewski
         }
\address {
                         Institute of Theoretical Physics,
                                Warsaw University,
                        Ho\.za 69, 00-681 Warsaw, Poland
         }
\maketitle

\begin{abstract}
We calculate positions of one- and two-particle, proton and
neutron drip lines within the Hartree-Fock-Bogoliubov theory
using Skyrme interaction.  We also determine an approximate
$r$-process path defined as a line where the neutron binding
energy is equal to 2~MeV.  A weakening of the nuclear shell
structure at drip lines is found and interpreted as resulting
from a coupling with continuum states.
\end{abstract}

\narrowtext

A description of nuclei far from the stability line is one of
the most important challenges for the nuclear structure theory.
Various methods of extrapolating measured nuclear masses to
large neutron or proton excess have already been proposed
\cite{AtDat}.  The proton-drip line has been reached and crossed
experimentally in several places of the nuclear chart.  On the
other hand, the neutron-drip line has been experimentaly riched
only for very light nuclei.  Its exact position for heavier
nuclides is still not known and theoretical predictions
\cite{AtDat} may differ by as much as 12 mass units for Z=50.

In the present study we report on calculations determining the
proton and neutron, one- and two-particle drip lines by using
the self-consistent mean-field approach. Since the position of
one-particle-drip lines depends crucially on a cancellation
between mean-field and pairing effects, the complete
Hartree-Fock-Bogoliubov (HFB) equations \cite{RS80} have to be
solved, which allow for a correct description of the interplay
between particle-particle and particle-hole channels of
interaction.  We use here the HFB theory with Skyrme interaction
\cite{DFT84}.

The HFB method has an advantage of giving nuclear density which
goes to zero at large distances, even for nuclei having Fermi
energy arbitrarily close but below zero.  At the same time, when
it is solved in the coordinate space it fully takes into account
the coupling of bound states to continuum \cite{DFT84}.  In this
way one avoids the appearance of unphysical particle gas which
would surround the nucleus if the BCS approximation had been
used to describe pairing correlations.  The HFB method has
recently been used \cite{HZD89} to describe properties of nuclei
far from stability constituting the neutron-star crust.

Many different parameterizations of the Skyrme force existing in
the literature have been obtained by fitting properties of known
nuclei.  Apart from standard versions \cite{BFG75,BQM82} used in
various nuclear structure calculations, there are recently
numerous attempts to obtain improved parameters.  A large-scale
adjustment of parameters to many nuclear masses has been
possible by using extended Thomas-Fermi approximation to the
Hartree-Fock (HF) method \cite{APD92} and the particle-drip
lines have been determined in this way.  A force devised for a
description of nuclei far from stability has been obtained
\cite{PR89} by fitting the Skyrme energy functional to the free
energy of nuclear and neutron matter calculated using
hyper-netted chain techniques.  Improved parameterizations have
been found in the seniority HF calculations \cite{GPN92}.  The
particle-drip lines have also been estimated from the
relativistic Hartree theory
\cite{HTW91}.

An extrapolation to nuclei far from stability can be meaningful
only if the force parameterization is used within the same
theoretical method wherewith it has been determined.
Unfortunately, the numerical effort of the HFB calculations is
large, and a large-scale fitting of parameters is still
unavailable.  Moreover, both the fitting and the extrapolation
should be made by using deformed HFB codes in the coordinate
space which do not yet exist.  In this study we present a pilot
calculation within the spherical approximation to the HFB method
as described in Ref.{\ }\cite{DFT84}, and we use the SkP force
parameterization, which has been obtained there by fitting
properties of several magic nuclei together with the Z=50
isotopic chain.

For fixed proton number $Z$, the neutron-drip line separates the
heaviest even-$N$ nucleus from the next odd-$N$ isotope such
that the neutron separation energy $S_n$ is negative,
\begin{equation}\label{e1}
S_n \equiv E^{\text{even}}(N)-E^{\text{odd}}(N+1) < 0 ,
\end{equation}
where $E^{\text{even}}(N)$ and $E^{\text{odd}}(N+1)$ denote
ground-state energies of neighbouring even and odd isotopes,
respectively.  In the HFB theory the number of particles is not
conserved and the energy is a function of the Lagrange
multiplier $\lambda$ called the Fermi energy.  By changing
$\lambda$ we may obtain HFB solutions with arbitrary (not
necessarily integer) average particle numbers.  Since the HFB
variational wave function contains only even-particle-number
components, the usual HFB solutions describe even nuclei even if
the average particle number $N$ is odd or not integer at all.
We denote the ground-state energies obtained in this way by
$\tilde{E}^{\text{even}}(\lambda)$.

In order to describe odd nuclei one should in principle use
variational function containing only odd-particle-number
components.  Usually one avoids this step by using the so-called
blocking approximation \cite{RS80}.  In the present study we
still simplify the calculations by approximating ground-state
energies of odd nuclei as
\begin{equation}\label{e2}
\tilde{E}^{\text{odd}}(\lambda) \simeq
   \tilde{E}^{\text{even}}(\lambda)+\min_{\mu}E_{\mu}(\lambda) ,
\end{equation}
where $E_{\mu}(\lambda)$ are the quasiparticle energies obtained
in the HFB theory for a given value of $\lambda$.  As shown in
Ref.  \cite{DFT84}, this is a fair approximation to the blocking
approach.

Because in the HFB theory the neutron number is a continuous
variable, we may fulfill the one-neutron-drip-line condition
(\ref{e1}) by looking for such two values of the Fermi energy
$\lambda$ and $\lambda'$ that
\begin{equation}\label{e3}
\tilde{E}^{\text{even}}(\lambda') -
       \tilde{E}^{\text{odd}}(\lambda) = 0
\end{equation}
and $N(\lambda')$=$N(\lambda) - 1$.  On the other hand, in the
HFB theory the Fermi energy is rigorously equal to the
derivative of ground-state energy with respect to the particle
number,
\begin{equation}
\lambda = \frac{\partial \tilde{E}^{\text{even}}(N)}{\partial N},
\end{equation}
and we may use the following first-order Taylor expansion:
\begin{equation}
\tilde{E}^{\text{even}}(N')
     \simeq \tilde{E}^{\text{even}}(N)
+ (N'-N)\frac{\partial \tilde{E}^{\text{even}}(N)}{\partial N} ,
\end{equation}
which gives
\begin{equation}\label{e4}
\tilde{E}^{\text{even}}(\lambda')
   \simeq \tilde{E}^{\text{even}}(\lambda) - \lambda .
\end{equation}
Inserting approximations (\ref{e2}) and (\ref{e4}) to the
one-neutron-drip-line condition (\ref{e3}) one obtains
\begin{equation}\label{e5}
       \lambda + \min_{\mu}E_{\mu}(\lambda) = 0 ,
\end{equation}
and this condition has been used in the calculations of the
present study.

The advantage of using such a condition consists in avoiding all
explicit calculations for odd nuclei as well as constraints for
neutron number N.  Once the HFB equations are solved under
condition (\ref{e5}) we can calculate the (not necessarily
integer) neutron number $N(\lambda)$ corresponding to the
obtained Fermi energy $\lambda$.  We then establish the
one-neutron-drip line as passing between the even isotope
closest to the obtained neutron number $N(\lambda)$ and the next
heavier odd isotope.  With this prescription and approximations
(\ref{e2}) and (\ref{e4}) one can determine the position of the
drip line with a precision of $\pm2$ mass units.

Outside the one-neutron-drip line all odd isotopes are unstable,
whereas the even ones are stable up to the two-neutron-drip line
defined by
\begin{equation}\label{e6}
S_{2n} \equiv E^{\text{even}}(N)-E^{\text{even}}(N+2) < 0 ,
\end{equation}
where $S_{2n}$ is the two-neutron separation energy, and
$E^{\text{even}}(N)$ and $E^{\text{even}}(N+2)$ denote
ground-state energies of neighbouring even isotopes,
respectively.  Considerations similar to those presented above
allow us to approximate condition (\ref{e6}) by the HFB
equations solved under the condition
\begin{equation}\label{e7}
\lambda=0 .
\end{equation}
This corresponds to simply omitting the neutron-number
constraint when solving the HFB equations.  Of course, the
constraint on proton number has always to be present when using
conditions (\ref{e5}) or (\ref{e7}) because it defines the
number of protons $Z$ at which we determine the positions of
neutron-drip lines.  In order to obtain the positions of
proton-drip lines we may repeat {\it mutatis mutandis} all above
considerations.

In our numerical calculations the neutron-drip lines have been
determined for even $Z$ between $Z$=8 and 124, and the
proton-drip lines for even $N$ between $N$=8 and 210, i.e., well
beyond the fission instability limits.  An overview of the
results is presented in Fig.{\ }\ref{fig1} where the scale is
large enough that one can read off the exact positions of drip
lines from the Figure.  The convention used is that of a nuclear
chart, i.e., the even-even nuclides are represented by squares
delimited by the pairs of tics on the abscissa and on the
ordinate.  Longer tics show limits of squares for particle
numbers divisible by 20.  The lines on the Figures separate
squares corresponding to stable and unstable nuclides.

The influence of closed major shells on proton-drip lines is
clearly visible at $Z$=50, 82, and 126.  At these proton numbers
the one- and two-proton-drip lines coincide and have long
horizontal sections centered around $N$=50, 104, and 194,
respectively.  This corresponds to closed proton shells to which
adding neither one nor two protons can produce a stable nucleus.
At large $Z$, the proton shell structure at proton-drip lines is
therefore the same as that for stable nuclei.  On the other
hand, for smaller proton numbers the shell structure is not
visible at proton drip lines.

Similar effects are seen in Fig.{\ }\ref{fig3} where we
present the average proton pairing gap $\Delta_p$ \cite{DFT84}
and the binding energy per particle $B/A$ calculated along the
one-proton-drip line as functions of the proton number.  At
large proton magic numbers $Z$=50, 82, and 126 the pairing
correlations disappear and the binding energy sharply increases.
At lower magic numbers the pairing correlations decrease but do
not vanish, while the binding energy has wide maxima, which is a
signature of a less pronounced shell structure. On the other
hand, at the one-neutron-drip line (i.e., when protons are
deeply bound) the proton pairing gap vanishes at every usual
magic number, Fig.{\ }\ref{fig3}.

A weakening of the shell structure at neutron-drip lines is
manifest when looking at the results presented in
Fig.{\ }\ref{fig1}.  Neither one- nor two-neutron-drip lines
have long vertical sections at constant $N$ values corresponding
to usual magic numbers.  Only wide bends are seen around
$N$=126, 184, and 258, which are the magic numbers of the
spherical shell model.  They illustrate the remaining influence
of the shell structure in nuclei with very large neutron excess.

This is confirmed by the behavior of the neutron pairing gap
$\Delta_N$ and of the binding energy along the neutron-drip
line, Fig.{\ }\ref{fig4}.  The weakness of a shell effect at
$N$=82 is especially striking.  It can probably be attributed to
a modified position of the 1h$_{11/2}$ orbital which at the
neutron-drip line is located in the middle of the $N$=82 shell
gap.  This supports the suggestion \cite{Pethick} that the
spin-orbit splitting may be smaller at large neutron excess,
because a larger surface diffuseness may lead to a decreased
strength of the spin-orbit interaction form-factor.  In fact, it
would also be very interesting to determine the spin-orbit
strength at large neutron excess, which can be done in the frame
of the relativistic Hartree-Fock theory \cite{Reinhard,Rufa},
and see whether it may modify the position of intruder states at
the neutron-drip line.

In Fig.{\ }\ref{fig1} we also present the results of the HFB
calculations with the neutron Fermi energy being fixed at the
value of 2~MeV.  This roughly corresponds to the approximate
$r$-process path \cite{CTT91}, which should pass through nuclei
which are at the origin of the abundance maxima.  Available
experimental information about the $r$-process path suggests
that it should pass through nuclides $^{80}_{30}$Zn$_{50}$,
$^{130}_{48}$Cd$_{82}$, and $^{195}_{69}$Tm$_{126}$ which in
Fig.{\ }\ref{fig1} are shown as full squares.

Our approximate $r$-process path goes through
$^{195}_{69}$Tm$_{126}$ at $N$=126, where a vertical section of
the path reappears.  This illustrates an increased role of the
shell structure when one moves away from the drip lines and when
the coupling with continuum states decreases.  On the other
hand, such vertical sections do not reappear at $N$=50 and 82,
and our $r$-process path misses the nuclides
$^{80}_{30}$Zn$_{50}$, $^{130}_{48}$Cd$_{82}$ by a few mass
units.  However, dynamic reaction networks $r$-process
calculations with the HFB masses should be performed if one
wants to properly assess the validity of the force parameters at
large neutron excess \cite{RoR88}.

In conclusion, the HFB calculations with the SkP Skyrme force
performed within the spherical approximation for nuclei with
large proton or neutron excess indicate that the shell structure
at magic particle numbers is weaker than that in stable nuclei.

We gratefully acknowledge helpful discussions with P.~Haensel
and A.~Sobiczewski.  This research was supported in part by the
Polish State Committee for Scientific Research under Contracts
No.  20450~91~01  and  No. 20954~91~01.

\widetext
\begin{figure}
\caption[F1]{One-particle (bottom) and two-particle (top)
neutron and proton drip lines obtained within the HFB theory
with the SkP Skyrme interaction. The middle line in the top part
of the Figure corresponds to an approximated $r$-process path.
Three full squares represent positions of the three nuclides
being at the origin of abundance maxima, i.e.,
$^{80}_{30}$Zn$_{50}$, $^{130}_{48}$Cd$_{82}$, and
$^{195}_{69}$Tm$_{126}$.}
\label{fig1}
\end{figure}
\narrowtext

\begin{figure}
\caption[F3]{Binding energy per particle $B/A$ (bottom) and the
proton pairing gap $\Delta_p$ (middle) calculated at the
one-proton-drip line as function of the proton number $Z$.  The
proton pairing gap $\Delta_p$ calculated at the one-neutron-drip
line is also shown (top).}
\label{fig3}
\end{figure}

\begin{figure}
\caption[F4]{Same as in Fig.{\ }\ref{fig3}, but for neutrons.}
\label{fig4}
\end{figure}

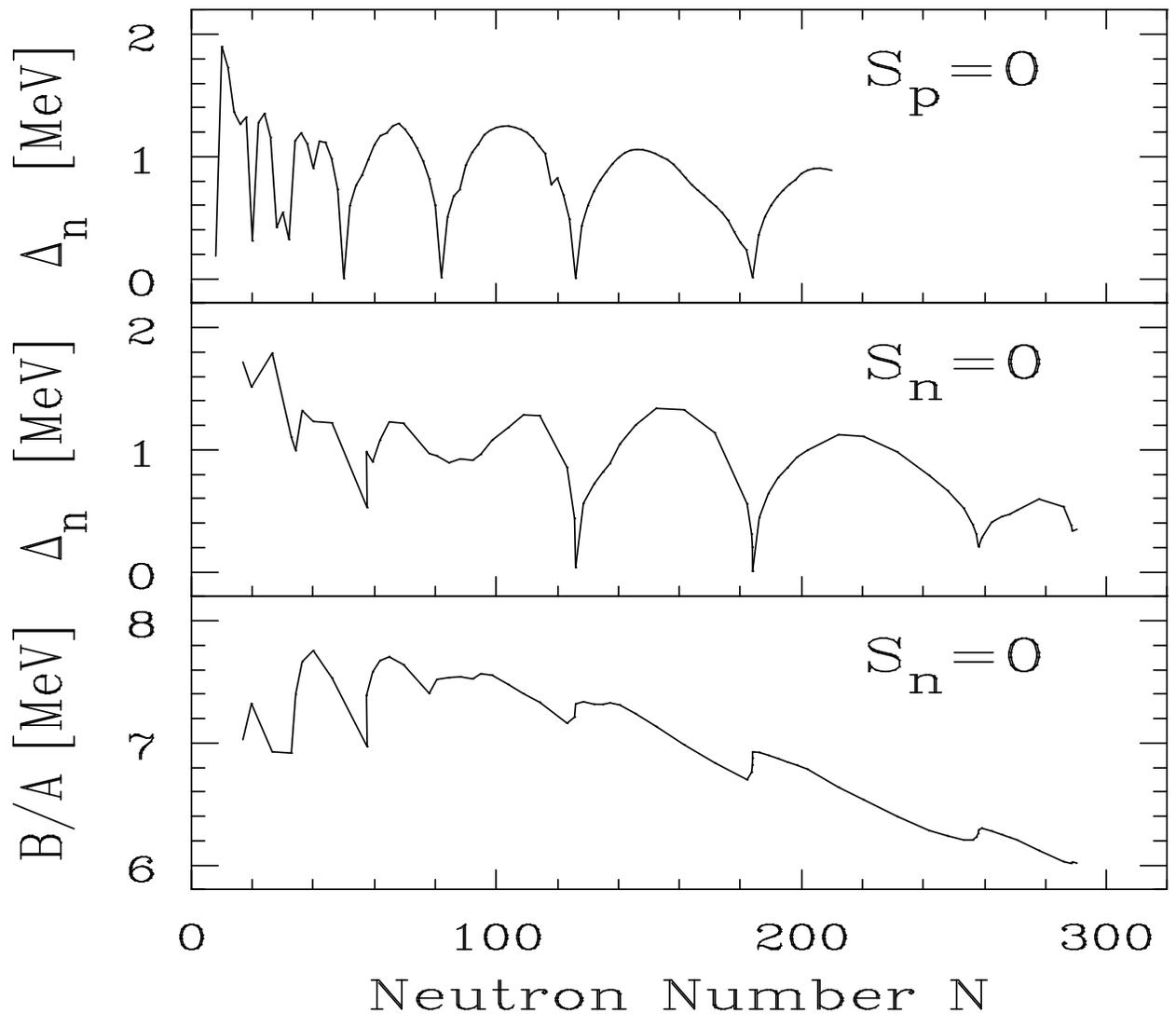
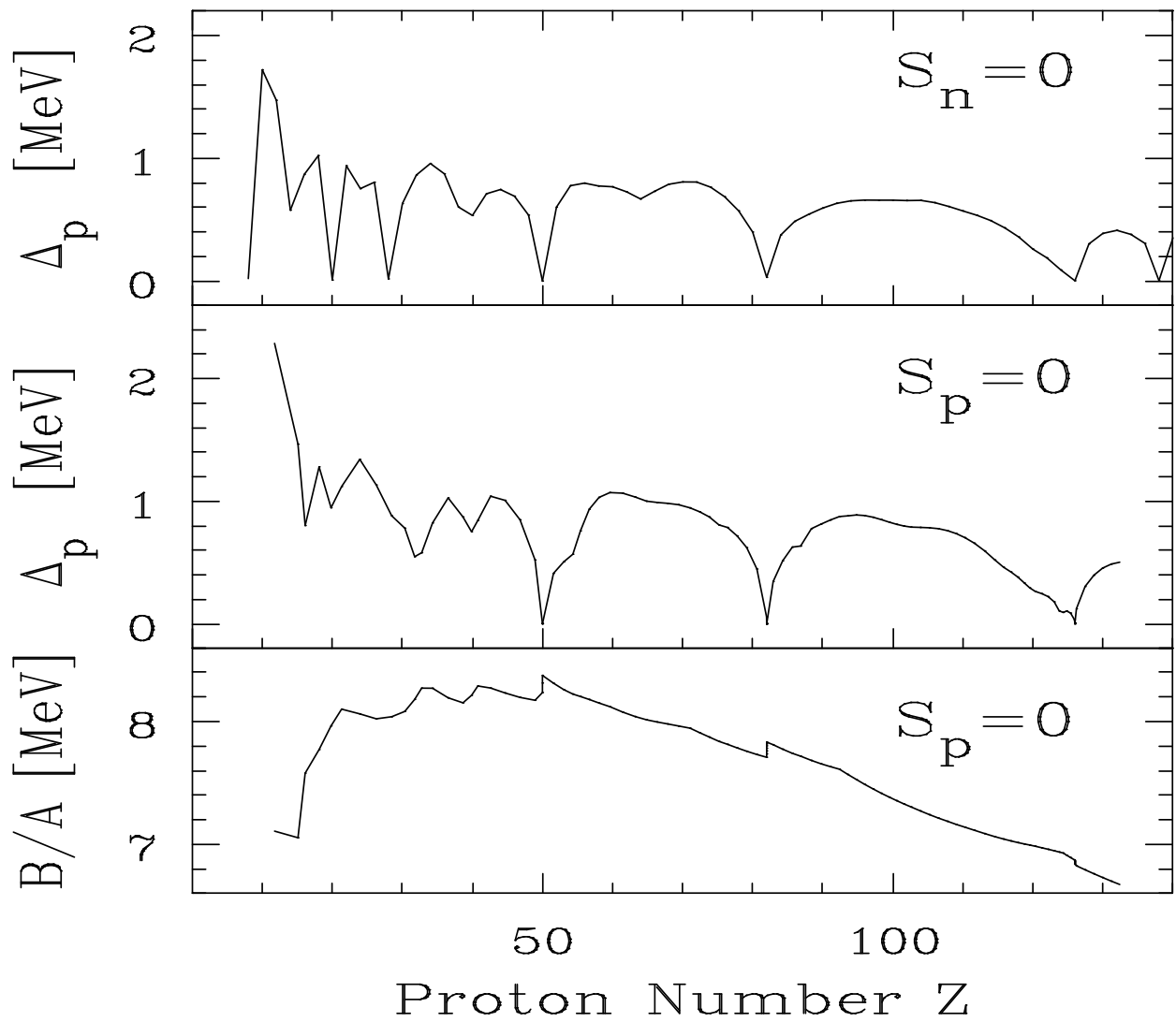
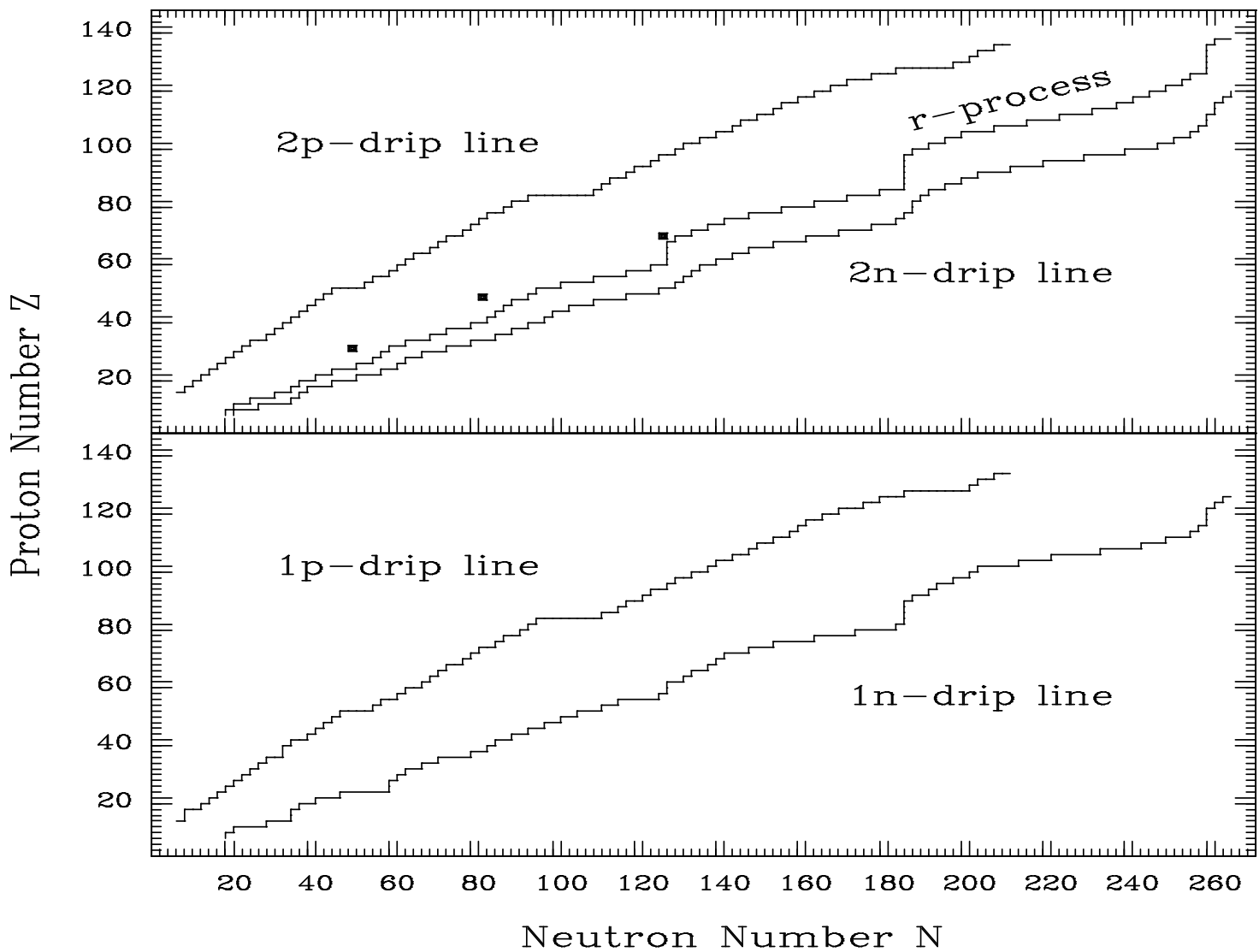
\end{document}